\documentclass{elsart}
\usepackage{graphicx}

\newcommand{\be}{\begin{equation}}
\newcommand{\ee}{\end{equation}}
\newcommand{\bea}{\begin{eqnarray}}
\newcommand{\eea}{\end{eqnarray}}

\newcommand{\BE}{\begin{eqnarray}}
\newcommand{\EE}{\end{eqnarray}}
\newcommand{\BEn}{\begin{eqnarray*}}
\newcommand{\EEn}{\end{eqnarray*}}
\newcommand{\barr}{\begin{array}}
\newcommand{\earr}{\end{array}}

\newcommand{\bit}{\begin{itemize}}      
\newcommand{\eit}{\end{itemize}}
\newcommand{\bc}{\begin{center}}
\newcommand{\ec}{\end{center}}
\newcommand{\ben}{\begin{enumerate}}    
\newcommand{\een}{\end{enumerate}}

\newcommand{\bR}{\ensuremath{\mathbf{R}}}
\newcommand{\bomega}{{\mbox{\boldmath $\omega$}}}
\newcommand{\bxi}{{\mbox{\boldmath $\xi$}}}
\begin{document}

\begin{frontmatter}
\title{The minority game: effects of strategy correlations and timing of adaptation}
\author{David Sherrington* and Tobias Galla}
\address{Theoretical Physics, 1 Keble Road, Oxford OX1 3NP, United
Kingdom 
\\
{\mdseries}* corresponding author, E-mail address: {\tt sherr@thphys.ox.ac.uk}}
\date{\today}


\begin{abstract}
A brief review is given of the minority game, an idealized model of a
market of speculative agents, and its complex many-body
behaviour. Particular consideration is given to the consequences and
implications of correlations between stategies and different
frequencies and timings of adaptation.
\end{abstract}
\end{frontmatter}
There is currently much interest in the statistical physics community
in the emergence of complex co-operative behaviour as a consequence of
frustration and disorder in systems of simple microscopic constituents
and rules of interaction. Examples are found in spin and structural
glasses, neural networks and hard optimization problems
\cite{N}. Minimal models, designed to capture the essence of real
world problems without the latter's peripheral complications or to
go beyond them to provide solubility and conceptual insight, have
played crucial roles in the understanding of such systems, via their
computer simulation, mean field analyses and extrapolation. The
minority game is a corresponding minimalist econophysics model
introduced to mimic a simple market of speculators trying to profit by
buying low and selling high. In this paper we review some of its
features and consider it in the context of complexity.

The model comprises an ensemble of a large number $N$ of agents each
of whom at each step of a discrete dynamics makes a choice which has a
scalar value that can be either positive or negative (buy or
sell). The objective of each agent is to make a choice of opposite
sign from that of the sum of all the actions. No agent has any direct
knowledge of the actions or propensities of the others but is aware of
the cumulative action (sum of the choices). Each agent's actual choice
is determined by the application of a personal strategy operator to
some common information available identically to all. In the simplest
versions of the model, to which we restrict here, the strategy
operators are chosen randomly and independently for each agent before
play commences and are not modified during play. Each agent does
however have a finite set of strategies, which for simplicity we
restrict to two each. At each step one strategy is chosen. The choice
is determined by points allocated to each strategy and augmented
regularly according to a comparison between the behaviour its play
would have yielded and the actual outcome; the points are increased
for minority prediction. This is the only mechanism for co-operation
but is sufficient to yield complex macroscopic behaviour. As we shall
see this complex behaviour has both similarities and differences as
compared to condensed matter systems studied earlier.

In the original version of the model \cite{CZ} the information on which
decisions were made was the history of the actual play over a finite
window (e.g. the last $m$ steps). However, simulations demonstrated that
utilising instead a random fictitous `history' at each time-step
produced essentially identical behaviour, suggesting that its
relevance is just to provide a mechanism for interaction \cite{Cav}.

Effective interaction via the common information is demonstrated by a
deviation of the macroscopic behaviour from that of agents making
independent random choices at each time-step. Phase transitions occur
as a function of scaled information dimension $d=D/N$, where $D$ is
the unscaled information dimension \cite{CZ2}, and also as a function
of `temperature' $T$ in the case of stochastic indeterminacy in
strategy choice \cite{CGGS,SCH}. These manifest in singularities in
macroscopic measures, such as the fraction of frozen agents \cite{CM}
and appropriate response functions \cite{HC}. Complexity arises in
that a critical phase line in $(d,T)$ space separates a region in
which the starting point allocations are irrelevant, after an
`equilibration' period, from one in which different starting extremes
yield different behaviours; we refer to these situations respectively
as {\em ergodic} and {\em non-ergodic}. The non-ergodic behaviour
manifests in the volatility being sensitive to initial conditions,
reminiscent of the differences between field-cooled and
zero-field-cooled linear susceptibilities in a spin glass
\cite{N}. While the volatility remains finite for all $(d,T)$, a
divergence does occur in an appropriate response function \cite{HC},
reminiscent of that in the spin-glass or third-order non-linear susceptibilities
in a spin glass.

Since the information on which the agents act is the same for all,
this problem is manifestly mean-field. It therefore offers the
potential for exact solution for its macro-behaviour in the sense of
the elimination of the microscopic variables in favour of
self-consistently determined macro-parameters in the thermodynamic
limit. The actual solution of these self-consistency equations is,
however, non-trivial in general, and not all scenarios have yet been
fully explored.

In this paper we restrict discussion to systems with the
`information' (on which the strategies operate to yield choices)
chosen randomly at each time-step. However we consider several
variations on further convention, in particular with respect to
independence of strategies and frequency or timing of the updating of
strategy points. These variations have demonstrated several novelties
some of whose investigation has thrown new light on the underlying
behaviour and others which require further study.  

The physics seems robust to variations of detail
concerning the use of continuous or discrete `bid' or strategy spaces,
nor on the number of strategies each agent holds provided it is
finite. However, for completeness we indicate the versions used in the
figures. At the basic level each agent $i,\, i=1,\dots,N$, is taken to
have two $D=dN$-dimensional strategies
$\bR_{ia}=(R_{ia}^1,\dots,R_{ia}^{dN}),\, a=\pm1$, with each component
$R_{ia}^\mu$ chosen independently randomly $\pm 1$ at the outset and
thereafter fixed. $\mu(t)$ is chosen stochastically randomly at each
time-step $t$ from the set $\mu(t)\in\{1,\dots,D\}$ and each agent
plays one of his or her two strategies $R_{ia}^{\mu(t)},\,a=\pm
1$. The actual choices of $a$ used, $b_i(t)$, are determined by the
current values of point differences $p_i(t)$. Here we restrict to
deterministic choices, $b_i(t)=\mbox{sgn}(p_i(t))$, but a simple
extension to stochastic choices is possible \cite{CGGS,SCH,CHS}. The $p_i(t)$
are updated every $M$ time-steps according to
\begin{equation}\label{onlineupdate}
\hspace{-0.5cm}
p_i(t+M)=p_i(t)-M^{-1}\sum_{\ell=t}^{t+M-1}
\xi_i^{\mu(\ell)}\left\{N^{-1/2}\sum_j(\omega_j^{\mu(\ell)}+\xi_j^{\mu(\ell)}\mbox{sgn}(p_j(t)))\right\},
\end{equation}
where
$\bomega_i=(\bR_{i1}+\bR_{i2})/2,\,\bxi_i=(\bR_{i1}-\bR_{i2})/2$. The
case where $M=1$ is referred to as the online model. For $M\geq O(N)$
one expects (and simulations confirm) the behaviour to be the same as
that for the so-called batch version in which the sum on the actual
$\mu(\ell)$ in (\ref{onlineupdate}) is replaced by an average:
\begin{equation}\label{batchupdate}
\hspace{-0.5cm}
p_i(t+1)=p_i(t)-\lambda N^{-1}\sum_{\mu=1}^{D}
\xi_i^{\mu}\left\{\sum_j(\omega_j^{\mu}+\xi_j^{\mu}\mbox{sgn}(p_j(t)))\right\},\,\lambda=O(1).
\end{equation}
\begin{figure}
\centerline{\includegraphics[width=33pc]{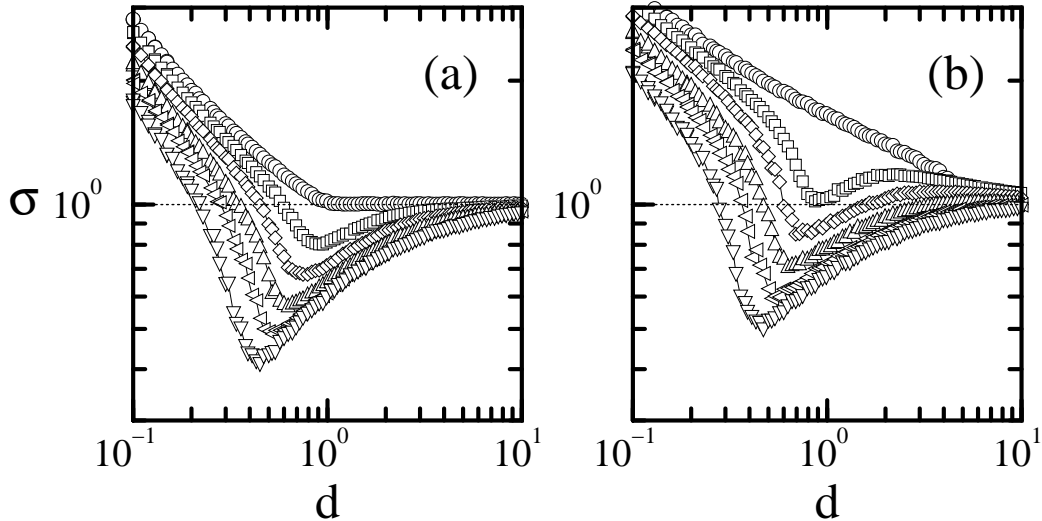}}
\caption{Volatilities with varying degrees of anticorrelation between
pairs of strategies, $P(R_{i1}^\mu=R_{i2}^\mu)\equiv\rho=0.0, 0.1,
0.2, 0.3, 0.4, 0.5$, top to bottom, (a) online, (b) batch.}
\label{figure1}
\end{figure}

It is of interest to compare the cases of differently correlated
$\bR_{i1,2}$ \cite{SGM}. In Fig. \ref{figure1} are shown the
volatilities obtained in simulations with strategies chosen randomly
with correlation probability $P(R_{i1}^\mu=R_{i2}^\mu)=\rho, \,
\rho\in[0,0.5]$, for both the on-line and the batch models. For the
uncorrelated case the online and batch models are almost identical,
both exhibiting a minimum in volatility lower than the random-bid
value at a critical $d$ \cite{CZ2,SMR}, but as the level of
(anti-)correlation is increased the volatilities of the online and
batch models become increasingly different from one another, as well
as different from the uncorrelated situation. Earlier studies on
systems with uncorrelated strategies showed slight differences between
online and batch models both in simulations \cite{GMS} and
theoretical studies \cite{CH} but anti-correlation as studied here
significantly amplifies and highlights the differences.

\begin{figure}
\centerline{\includegraphics[width=33pc]{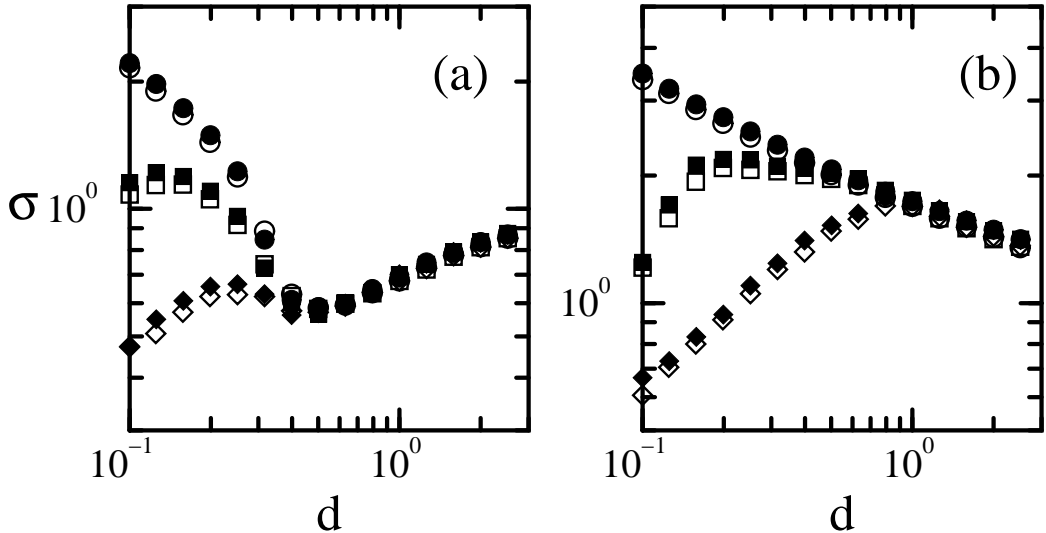}}
\caption{Comparison of volatilities for {\em tabula rasa} and strongly
biased-point starts in batch minority games after $40$ time-steps
($p_i(0)=0.0$ (circles), $0.5$ (squares) and $1.0$ (diamonds) for
normalisation $\lambda=(1-\rho)^{-1}$). (a) Uncorrelated
strategies. (b) Fully anti-correlated case. Open symbols denote
simulations for $N=1000$ players. Solid symbols represent results from
a numerical evaluation of the macroscopic self-consistent equations
derived analytically in the limit $N\to\infty$. An average is
performed over 50000 realisations of the single-particle noise.}
\label{figure2}
\end{figure}

In the thermodynamic limit ($N\to\infty$) the macrodynamics of the batch
problem for arbitrarily correlated local strategies $\bR_{i1,2}$ can
be expressed in terms of non-linear coupled equations involving
two-time correlation and response functions, via an extension
of the generating functional method of Heimel and Coolen
\cite{HC}. The full equations are complicated but can be conveniently
expressed via a self-consistently determined coloured effective
single-particle noise \cite{HC,GS}. Here we just give some
results. Let us concentrate first on the case of independent
$\bR_{i1,2}$. In this case the volatility exhibits a minimum at a
critical $d_c$ above which the dynamics is ergodic and beneath which
it is non-ergodic, the volatility yielded by a {\em tabula rasa} point
initiation, $p_i(0)=0$, being greater than that for a strongly biased
start $|p_i(0)|\sim O(1)$. For $d\geq d_c$ a simple ansatz provides the
stationary limit in accord with simulations \cite{SMR}. For $d<d_c$ the
behaviour is non-ergodic and a full closed analytic solution is still
awaited, but iteration of the coupled macrodynamical equations is
possible numerically using the self-consistent noise-sampling
procedure of Eissfeller and Opper \cite{EO}. This shows good accord
with corresponding simulations, as is illustrated in
Fig. \ref{figure2}a. An analagous treatment of the anti-correlated
case demonstrates that the non-ergodic transition persists, even
though the volatility itself appears smooth; see
Fig. \ref{figure2}b. The existence of a phase transition separating
ergodic and non-ergodic regions is reminiscent of transitions between
equilibrating and non-equilibrating regions in the `condensed matter'
examples mentioned earlier. The existence of a critical dimension
$D_c$ scaling as $N$ is reminiscent of the critical scaling limit on
the number of retrievable stored patterns in a neural network
\cite{H}, but there are also crucial differences \cite{CMZ,SGM}.

For $M\geq O(N)$ the temporal correlation functions 
\begin{equation}
C(\tau)= \lim_{t\to\infty}
N^{-1}\sum_i\mbox{sgn}(p_i(t+\tau))\mbox{sgn}(p_i(t))
\end{equation}
\vspace{-1.5em}
exhibit oscillations with period $2M$. In the ergodic region these
oscillations die away with a finite decay time for $\rho>0$. In the non-ergodic
region oscillatory behaviour persists. Randomizing the precise time of
updating, independently for each agent, but maintaining the average
updating frequency, reduces and eventually (if the time-spread is
sufficient) removes the oscillations. This does not have an obvious
well-known spin glass analogue and suggests a different complexity.

The minority game was already recognized as having interesting
features from the perspective of complex emergent behaviour in many
body physics, including an ergodic-nonergodic phase transition. By
considering correlated strategies we have re-iterated that
anti-correlation increases volatility and have demonstrated that it
amplifies differences between online and batch behaviour, implying
that updating frequency is relevant in generalizations of online
learning models. We have also shown that analytic theory of
the batch model can be extended to correlated strategies and is in good
accord with simulations for both ergodic and
non-ergodic regions. An ergodic-nonergodic transition is found in the
fully anti-correlated case even though the {\em tabula rasa}
volatility is smooth.

A more complete description of our results and analysis will be presented 
elsewhere \cite{GS}.

{\bf Acknowledgements}

The authors would like to thank the EPSRC for financial support under
research grant GR/M04426 and studentship 00309273……. TG 
acknowledges the award of a Rhodes Scholarship. Both would like to
thank colleagues for valuable discussions, especially A.C.C.Coolen and
J.P.Garrahan.

\end{document}